\documentclass[10pt]{article}
\usepackage[totalwidth=450pt, totalheight=590pt, centering]{geometry}

\usepackage{amsmath,amssymb,mathrsfs}
\usepackage{bbm} 

\usepackage[dvips]{graphicx}
\usepackage{subfigure}

\usepackage[dvips,bookmarks=true,hyperfootnotes=false]{hyperref} 
\hypersetup{pdfstartview=FitH,pdfhighlight=/O,colorlinks=false}

\newcommand{\mc}[1]{\mathcal{#1}}

\newcommand{\de}{\delta}

\newcommand{\mdots}{,.\,.\,,}
\newcommand{\onehalf}{\frac{1}{2}}
\long\def\symbolfootnote[#1]#2{\begingroup\def\thefootnote{\fnsymbol{footnote}}
\footnote[#1]{#2}\endgroup}

\begin{document}
\title{\Large\bf Asymptotics of LQG fusion coefficients}

\author{Emanuele Alesci {\it ${}^{1ab}$}, Eugenio Bianchi {\it ${}^{2ac}$}, Elena Magliaro {\it ${}^{3ad}$}, Claudio Perini {\it ${}^{4ae}$}\\[.35em]
\small{\textit{${}^a$Centre de Physique Th\'eorique de Luminy}}\footnote{Unit\'e mixte de recherche (UMR 6207) du CNRS et des Universit\'es
de Provence (Aix-Marseille I), de la M\'editerran\'ee (Aix-Marseille II) et du Sud (Toulon-Var); laboratoire affili\'e \`a la FRUMAM (FR 2291).}\small{\textit{, case 907, F-13288 Marseille, EU}}\\
\small{\textit{${}^b$Laboratoire de Physique, ENS Lyon, CNRS UMR 5672, F-69007 Lyon, EU}}\\
\small{\textit{${}^c$Scuola  Normale Superiore, Piazza dei Cavalieri 7, I-56126 Pisa,  EU}}\\
\small{\textit{${}^d$Dipartimento di Fisica, Universit\`a degli Studi Roma Tre, I-00146 Roma, EU}}\\
\small{\textit{${}^e$Dipartimento di Matematica, Universit\`a degli Studi Roma Tre, I-00146 Roma, EU}}
}

\date{\small\today} \maketitle

\symbolfootnote[0]{e-mail: ${}^{1}$alesci@fis.uniroma3.it, ${}^{2}$e.bianchi@sns.it, ${}^{3}$elena.magliaro@gmail.com, ${}^{4}$claude.perin@libero.it}

\begin{abstract}
The fusion coefficients from $SO(3)$ to $SO(4)$ play a key role in the definition of spin foam models for the dynamics in Loop Quantum Gravity. In this paper we give a simple analytic formula of the EPRL fusion coefficients. We study the large spin asymptotics and show that they map $SO(3)$ semiclassical intertwiners into $SU(2)_L\times SU(2)_R$ semiclassical intertwiners.
This non-trivial property opens the possibility for an analysis of the semiclassical behavior of the model. 
\end{abstract}

\section{Introduction}
The recent construction of a class of spinfoam models \cite{Engle:2007uq,Engle:2007qf,Engle:2007wy,Freidel:2007py,Livine:2007vk} compatible with loop quantum gravity (LQG) \cite{Rovelli:1993gcqft,Rovelli:2004tv,Ashtekar:2004eh,Thiemann:2007zz} has opened the possibility of consistently defining the LQG dynamics using spinfoam techniques \cite{Baez:1997zt,Oriti:2001qu,Perez:2003vx,Alesci:2008yf}. In this paper we focus on the Engle-Pereira-Rovelli-Livine (EPRL) spinfoam model for Riemannian gravity introduced in \cite{Engle:2007wy}. For given Immirzi parameter $\gamma$, the vertex amplitude is defined as follows: it is a function of five $SO(3)$ intertwiners $i_a$ and ten spins $j_{ab}$ (with $a,b=1\mdots 5$ and $a<b$) given by
\begin{equation}
	W(j_{ab},i_a)=\sum_{i_a^L\,i_a^R}\, \{15j\}_N\big(\frac{|1-\gamma|j_{ab}}{2},i_a^L\big)\;\, \{15j\}_N\big(\frac{(1+\gamma)j_{ab}}{2},i_a^R\big)\;
	\prod_a  f^{i_a}_{i_a^L i_a^R}(j_{ab})\;.
	\label{eq:EPRL}
\end{equation}
The functions $\{15j\}_N$ are normalized $15j$-symbols, namely the contraction of five normalized 4-valent $SU(2)$ intertwiners according to the pattern of a 4-simplex, and the $f^{i_a}_{i_a^L i_a^R}$ are fusion coefficients from $SO(3)$ to $SU(2)_L\times SU(2)_R$ introduced in \cite{Engle:2007wy} and defined below. Such coefficients play a key role in the definition of the model. Indeed the model differs from the one introduced by Barrett and Crane \cite{Barrett:1997gw} only for the structure of these coefficients. In this paper we study the large spin asymptotics of the EPRL fusion coefficients.

A careful analysis of the asymptotics of fusion coefficients is a step needed for the study of the semiclassical properties of the model. In fact, we have already used the results that we present in this paper in order to understand the features of the wavepacket evolution. The propagation of semiclassical wavepackets was introduced in \cite{Magliaro:2007ni} as a new way to test the semiclassical limit of a spinfoam model. A spinfoam model has a good semiclassical behavior if semiclassical wavepackets (peaked on a classical 3-geometry) follow the trajectories predicted by the classical equations of motion. In \cite{Magliaro:2007ni} this new technique was implemented in the EPR flipped vertex model to study the propagation of intertwiner wavepackets. In \cite{abmp:2008id} we developed a more efficient numerical algorithm (using techniques similar to \cite{Khavkine:2008kk}) and applied the asymptotic analysis presented here. This kind of study is, in the general context of spinfoam models, complementary to the semiclassical analysis based on the calculation of $n$-point functions. 

In \cite{Rovelli:2005yj,Bianchi:2006uf}, a strategy for recovering graviton correlations from a background-independent theory was introduced. The idea was tested on the Barrett-Crane model at the ``single-vertex'' level. At this level, correlations of geometric operators can be checked against perturbative Regge-calculus with a single $4$-simplex \cite{Bianchi:2007vf}. Given the fact that the Barrett-Crane model gives trivial dynamics to intertwiners, the analysis was restricted to the spin degrees of freedom -- namely to area correlations only. On the other hand, the new models are consistent with the LQG kinematics and allow the computations of semiclassical correlations of geometric observables as the area, the angle, the volume or the length \cite{Rovelli:1994ge,Ashtekar:1996eg,Ashtekar:1997fb,Major:1999mc,Thiemann:1996at,Bianchi:2008es}. At the single-vertex level, the semiclassical correlations for two local geometric operators $\hat{\mc{O}}_1$, $\hat{\mc{O}}_2$ are simply given by 
\begin{equation}
	\langle\hat{\mc{O}}_1\, \hat{\mc{O}}_2 \rangle_q=\frac{\sum_{j_{ab} i_a} W(j_{ab},i_a)\; \hat{\mc{O}}_1\, \hat{\mc{O}}_2\;\Psi_q(j_{ab},i_a)}{\sum_{j_{ab} i_a} W(j_{ab},i_a)\,\Psi_q(j_{ab},i_a)}\;,
	\label{eq:correlation}
\end{equation}
where $W(j_{ab},i_a)$ is the vertex-amplitude introduced in (\ref{eq:EPRL}) and $\Psi_q(j_{ab},i_a)$ is a boundary semiclassical state peaked on a configuration $q$ of the intrinsic and the extrinsic geometry of the boundary of a region of space-time. The appropriate dependence on spins and intertwiners of the state $\Psi_q(j_{ab},i_a)$ is discussed in \cite{Alesci:2007tx,Alesci:2007tg} and uses the semiclassical tetrahedron state of \cite{Rovelli:2006fw}. Moreover, in order to guarantee that the appropriate correlations are present, in \cite{Alesci:2007tx,Alesci:2007tg} a specific form of the large spin asymptotics for the vertex amplitude was conjectured (see \cite{Alesci:2008gv}). In order to show that the EPRL vertex amplitude satisfies this conjecture, an analysis of the asymptotics of the fusion coefficients is needed. The region of parameter space of interest is large spins $j_{ab}$ and intertwiners $i_a$ of the same order of magnitude of the spins. As a result, the fusion coefficients for the node $\bar{a}$,  $f^{i_{\bar{a}}}_{i_{\bar{a}}^L i_{\bar{a}}^R}(j_{\bar{a} b})$, can be seen as a function of the two bare variables $i_{\bar{a}}^L$, $i_{\bar{a}}^R$, of the fluctuation of the intertwiner $i_{\bar{a}}$ and of the fluctuation of the four spins $j_{\bar{a}b}$. In this paper we focus on this analysis. For different approaches to the semiclassical limit, see \cite{Bianchi:2008ae} and \cite{Conrady:2008mk}.
 
The paper is organized as follows: in section \ref{sec:analytic expr} we show a simple analytic expression for the EPRL fusion coefficients; in section \ref{sec:asymptotics} we use this expression for the analysis of the asymptotics of the coefficients in the region of parameter space of interest; in section \ref{sec:semiclassical} we show that the fusion coefficients map $SO(3)$ semiclassical intertwiners into $SU(2)_L\times SU(2)_R$ semiclassical intertwiners. We conclude discussing the relevance of this result for the analysis of the semiclassical behavior of the model. In the appendix we collect some useful formula involving Wigner coefficients.

\section{Analytical expression for the fusion coefficients}\label{sec:analytic expr}
The fusion coefficients provide a map from four-valent $SO(3)$ intertwiners to four-valent $SO(4)$ intertwiners. They can be defined in terms of contractions of $SU(2)$ $3j$-symbols. In the following we use a planar diagrammatic notation for $SU(2)$ recoupling theory \cite{YutsinLevinson:1962}. We represent the $SU(2)$ Wigner metric  and  the $SU(2)$ three-valent intertwiner respectively by an oriented line and by a node with three links oriented counter-clockwise\footnote{A minus sign in place of the $+$ will be used to indicate clockwise orientation of the links.}. A four-valent $SO(3)$ intertwiner $|i\rangle$ can be represented in terms of the recoupling basis as \begin{equation}
	|i\,\rangle=\sqrt{2i+1} \parbox[c]{100pt}{\includegraphics{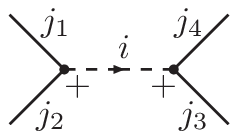}}
	\label{eq:feyn.i}
\end{equation}
where a dashed line has been used to denote the virtual link associated to the coupling channel. Similarly a four-valent $SO(4)$ intertwiner can be represented in terms of an $SU(2)_L\times SU(2)_R$ basis as $|i_L\rangle|i_R\rangle$. 

Using this diagrammatic notation, the EPRL fusion coefficients for given Immirzi parameter $\gamma$ are given by
\begin{align}\label{eq:f=feyn}
	f^i_{i_L i_R}(j_1,j_2,j_3,j_4) =&(-1)^{j_1-j_2+j_3-j_4}\sqrt{(2i+1)(2i_L+1)(2i_R+1)\Pi_{n=1}^4(2j_n+1)}\quad\times\\\nonumber
	  &\times\quad\parbox[c]{150pt}{\includegraphics{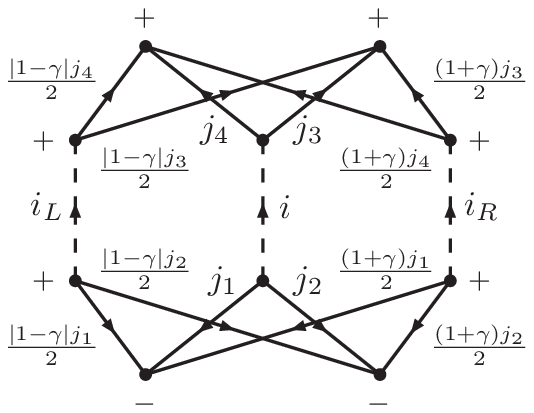}}\quad\;.
	\end{align}
These coefficients define a map
\begin{align}
f:Inv[H_{j_1}\otimes\ldots\otimes H_{j_4}]\longrightarrow Inv[H_{(\frac{|1-\gamma|j_1}{2},\frac{(1+\gamma) j_1}{2})}\otimes\ldots\otimes H_{(\frac{|1-\gamma|j_4}{2},\frac{(1+\gamma) j_4}{2})}]
\label{eq:fmap}
\end{align}
from $SO(3)$ to $SO(4)$ intertwiners. Using the identity
\begin{align}
&\parbox[c]{80pt}{\includegraphics{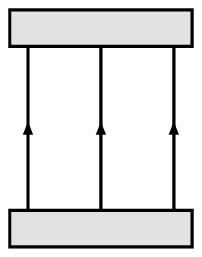}} =\;\; \parbox[c]{50pt}{\includegraphics{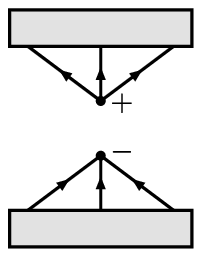}} 	\label{eq:3lines}\\[-1em]
& \nonumber
\end{align}
where the shaded rectangles represent arbitrary closed graphs, we have that the diagram in (\ref{eq:f=feyn}) can be written as the product of two terms
\begin{equation}
	f^i_{i_L i_R}(j_1,j_2,j_3,j_4) = (-1)^{j_1-j_2+j_3-j_4}\sqrt{(2i+1)(2i_L+1)(2i_R+1)\Pi_{n}(2j_n+1)}\;\; q^i_{i_L i_R}(j_1,j_2)\; q^i_{i_L i_R}(j_3,j_4)
	\label{eq:f=qq}
\end{equation}
where $q^i_{i_L i_R}$ is given by the following $9j$-symbol
\begin{equation}
	q^i_{i_L i_R}(j_1,j_2)\quad=\quad \parbox[c]{150pt}{\includegraphics{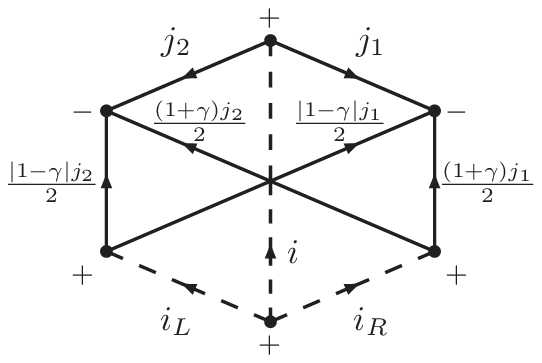}} \quad = \left\{\!\!
	\begin{array}{ccc}
  \frac{|1-\gamma|}{2} j_1 & i_L & \frac{|1-\gamma|}{2} j_2\\[3pt]
  \frac{1+\gamma}{2} j_1& i_R & \frac{1+\gamma}{2} j_2\\[3pt]
  j_1 & i & j_2\\
  \end{array}
	\!\!\right\}\;.
	\label{eq:q=feyn}
\end{equation}
From the form of $q^i_{i_L i_R}$ we can read a number of properties of the fusion coefficients. First of all, the diagram in expression (\ref{eq:q=feyn}) displays a node with three links labelled $i, i_L, i_R$. This corresponds to a triangular inequality between the intertwiners $i, i_L, i_R$ which is not evident from formula (\ref{eq:f=feyn}). As a result we have that the fusion coefficients vanish outside the domain 
\begin{equation}
	|i_L-i_R|\leq i \leq i_L+i_R\;.
	\label{eq:domain}
\end{equation}
Moreover in the monochromatic case, $j_1=j_2=j_3=j_4$, we have that the fusion coefficients are non-negative (as follows from (\ref{eq:f=qq})) and, for $i_L+i_R+i$ odd, they vanish (because the first and the third column in the $9j$-symbol are identical). 

As discussed in \cite{Freidel:2007py,Livine:2007vk}, the fact that the spins labeling the links in (\ref{eq:f=feyn}) have to be half-integers imposes a quantization condition on the Immirzi parameter $\gamma$. In particular $\gamma$ has to be rational and a restriction on spins may be present. Such restrictions are absent in the Lorentzian case. Now notice that for $0\leq\gamma<1$ we have that $\frac{1+\gamma}{2}+\frac{|1-\gamma|}{2}=1$, while for $\gamma>1$ we have that $\frac{1+\gamma}{2}-\frac{|1-\gamma|}{2}=1$ (with the limiting case $\gamma=1$ corresponding to a selfdual connection). As a result, in the first and the third column of the $9j$-symbol in (\ref{eq:q=feyn}), the third entry is either the sum or the difference of the first two. In both cases the $9j$-symbol admits a simple expression in terms of a product of factorials and of a $3j$-symbol (see appendix \ref{app:A}). Using this result we have that, for $0\leq\gamma<1$, the coefficient $q^i_{i_L i_R}(j_1,j_2)$ can be written as
\begin{equation}
q^i_{i_L i_R}(j_1,j_2) =	 (-1)^{i_L-i_R+(j_1-j_2)} \left(\!\!
\begin{array}{ccc}
i_L & i_R & i \\[4pt]
\frac{|1-\gamma|(j_1-j_2)}{2}& \frac{(1+\gamma)(j_1-j_2)}{2}  & -(j_1-j_2)	
\end{array}
\!\!\right)\;\; A^i_{i_L i_R}(j_1,j_2)
	\label{eq:q=analytic}
\end{equation}
with $A^i_{i_L i_R}(j_1,j_2)$ given by
\begin{align}
	A^i_{i_L i_R}(j_1,j_2) &=
	\;\; \sqrt{\frac{(j_1+j_2-i)!\, (j_1+j_2+i+1)!}{(2j_1+1)!\,(2j_2+1)!}}\;\;\; \times 	\label{eq:A=analytic} \\[6pt]
	&\;\times\; \sqrt{\frac{(|1-\gamma| j_1)!\, (|1-\gamma| j_2)!}{    \big(\frac{|1-\gamma|j_1}{2} + \frac{|1-\gamma|j_2}{2} -i_L\big)!\, \big(\frac{|1-\gamma|j_1}{2} + \frac{|1-\gamma|j_2}{2} +i_L+1\big)!}}\;\;\; \times\nonumber\\[6pt]
	&\;\times\; \sqrt{\frac{((1+\gamma)j_1)!\, ((1+\gamma)j_2)!}{ \big(\frac{(1+\gamma)j_1}{2} + \frac{(1+\gamma)j_2}{2} -i_R\big)!\, \big(\frac{(1+\gamma)j_1}{2} + \frac{(1+\gamma)j_2}{2} +i_R+1\big)!	}}\;\;.\nonumber
\end{align}
A similar result is available for $\gamma>1$. The Wigner $3j$-symbol in expression (\ref{eq:q=analytic}) displays explicitly the triangle inequality (\ref{eq:domain}) among the intertwiners. Notice that the expression simplifies further in the monochromatic case as we have a $3j$-symbol with vanishing magnetic indices.

The fact that the fusion coefficients (\ref{eq:f=feyn}) admit an analytic expression which is so simple is certainly remarkable. The algebraic expression (\ref{eq:f=qq}),(\ref{eq:q=analytic}),(\ref{eq:A=analytic}) involves no sum over magnetic indices. On the other hand, expression (\ref{eq:f=feyn}) involves ten $3j$-symbols (one for each node in the graph) and naively fifteen sums over magnetic indices (one for each link). In the following we will use this expression as starting point for our asymptotic analysis.

\section{Asymptotic analysis}\label{sec:asymptotics}
The new analytic formula (\ref{eq:f=qq}),(\ref{eq:q=analytic}),(\ref{eq:A=analytic}) is well suited for studying the behavior of the EPRL fusion coefficients in different asymptotic regions of parameter space. In this paper we focus on the region of interest in the analysis of semiclassical correlations as discussed in the introduction. This region is identified as follows: let us introduce a large spin $j_0$ and a large intertwiner (i.e. virtual spin in a coupling channel) $i_0$; let us also fix the ratio between $i_0$ and $j_0$ to be of order one -- in particular we will take $i_0=\frac{2}{\sqrt{3}}j_0$; then we assume that
\begin{itemize}
	\item the spins $j_1$, $j_2$, $j_3$, $j_4$, are restricted to be of the form $j_e=j_0+\de j_e$ with the fluctuation $\de j_e$ small with respect to the background value $j_0$. More precisely we require that the relative fluctuation $\frac{\de j_e}{j_0}$ is of order $o(1/\sqrt{j_0})$;
	\item the $SO(3)$ intertwiner $i$ is restricted to be of the form $i=i_0+\de i$ with the relative fluctuation $\frac{\de i}{i_0}$ of order $o(1/\sqrt{j_0})$;
	\item the intertwiners for $SU(2)_L$ and $SU(2)_R$ are studied in the region close to the background values $i^0_L=\frac{|1-\gamma|}{2} i_0$ and $i^0_R=\frac{1+\gamma}{2} i_0$. We study the dependence of the fusion coefficients on the fluctuations of these background values assuming that the relative fluctuations $\de i_L/i_0$ and $\de i_R/i_0$ are of order $o(1/\sqrt{j_0})$.
\end{itemize}
A detailed motivation for these assumptions is provided in section \ref{sec:semiclassical}. Here we notice that, both for $0\leq\gamma< 1$ and for $\gamma>1$, the background value of the intertwiners $i_L$, $i_R$, $i$, saturate one of the two triangular inequalities (\ref{eq:domain}). As a result, we have that the fusion coefficients vanish unless the perturbations on the background satisfy the following inequality
\begin{align}
	 &\de i\leq \de i_L + \de i_R \quad\quad 0\leq\gamma<1\\
	    &\de i_R\leq \de i + \de i_L \quad\quad\quad \gamma>1\;.
	\label{eq:de i inequality}
\end{align}
In order to derive the asymptotics of the EPRL fusion coefficients in this region of parameter space we need to analyze both the asymptotics of the $3j$-symbol in (\ref{eq:q=analytic}) and of the coefficients $A^i_{i_L i_R}(j_1,j_2)$. This is done in the following two subsections.

\subsection{Asymptotics of $3j$-symbols}
The behavior of the $3j$-symbol appearing in equation (\ref{eq:q=analytic}) in the asymptotic region described above is given by Ponzano-Regge asymptotic expression (equation 2.6 in \cite{PonzanoRegge:1968}; see also appendix \ref{app:B}):

\begin{align}
&\left(\!\!
\begin{array}{ccc}
i_L & i_R & i \\[4pt]
\frac{|1-\gamma|(j_1-j_2)}{2}& \frac{(1+\gamma)(j_1-j_2)}{2}  & -(j_1-j_2)	
\end{array}
\!\!\right)
 \sim \\[6pt]
& \quad\sim\frac{(-1)^{i_L+i_R-i+1}}{\sqrt{2 \pi A}} \cos\Big( (i_L+\frac{1}{2})\theta_L+(i_R+\frac{1}{2})\theta_R+(i+ \frac{1}{2})\theta+ {\textstyle \frac{|1-\gamma|(j_1-j_2)}{2}}  \phi_- - {\textstyle\frac{(1+\gamma)(j_1-j_2)}{2} }\phi_+ +\frac{\pi}{4}\Big)\;.\nonumber
 	\label{eq:3jPR}
\end{align}
The quantities $A$, $\theta_L$, $\theta_R$, $\theta$, $\phi_-$, $\phi_+$ admit a simple geometrical representation: let us consider a triangle with sides of length $i_L+\frac{1}{2}$, $i_R+\frac{1}{2}$, $i+\frac{1}{2}$ embedded in 3d Euclidean space as shown below
\begin{equation}
	\parbox[c]{300pt}{\includegraphics{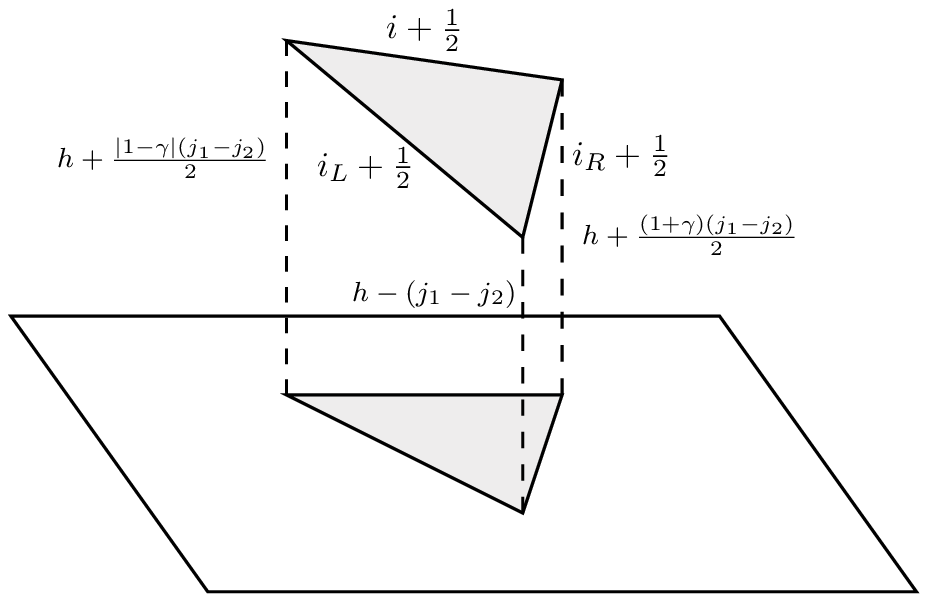}}
	\label{eq:}
\end{equation}
In the figure the height of the three vertices of the triangle with respect to a plane are given; this fixes the orientation of the triangle and forms an orthogonal prism with triangular base. The quantity $A$ is the area of the base of the prism (shaded in picture). The quantities $\theta_L$, $\theta_R$, $\theta$ are dihedral angles between the faces of the prism which intersect at the sides $i_L$, $i_R$, $i$ of the triangle. The quantities $\phi_-$, $\phi_+$ are dihedral angles between the faces of the prism which share the side of length $h+|1-\gamma|(j_1-j_2)/2$ and the side of length $h+(1+\gamma)(j_1-j_2)/2$, respectively. For explicit expressions we refer to the appendix.

In the monochromatic case, $j_1=j_2$, we have that the triangle is parallel to the plane and the formula simplifies a lot; in particular we have that the area $A$ of the base of the prism is simply given by Heron formula in terms of $i_L$, $i_R$, $i$ only, and the dihedral angles $\theta_L$, $\theta_R$, $\theta$ are all equal to $\pi/2$. As a result the asymptotics is given by
\begin{equation}
	\left(\!\!
\begin{array}{ccc}
i_L & i_R & i \\[4pt]
0 & 0  & 0	
\end{array}
\!\!\right)
 \sim 
 \frac{1}{\sqrt{2 \pi A}}\frac{1+(-1)^{i_L + i_R + i}}{2} (-1)^{\frac{i_L + i_R + i}{2}}\;.
	\label{eq:3j0}
\end{equation}
Notice that the sum $i_L + i_R + i$ is required to be integer and that the asymptotic expression vanishes if the sum is odd and is real if the sum is even. Now, the background configuration of $i_L$, $i_R$ and $i$ we are interested in corresponds to a triangle which is close to be degenerate to a segment. This is due to the fact that $\frac{(1-\gamma)}{2} i_0+\frac{(1+\gamma)}{2} i_0=i_0$ for $0\leq\gamma<1$, and $\frac{(\gamma+1)}{2} i_0-\frac{(\gamma-1)}{2} i_0=i_0$ for $\gamma>1$. In fact the triangle is not degenerate as an offset $\frac{1}{2}$ is present in the length of its edges. As a result the area of this almost-degenerate triangle is non-zero and scales as $i_0^{3/2}$ for large $i_0$. When we take into account allowed perturbations of the edge-lengths of the triangle we find
\begin{align}
	A=\begin{cases}
	\frac{1}{4} \sqrt{1-\gamma^2}\; i_0^{3/2}\,\big(\sqrt{1+2(\de i_L+\de i_R-\de i)}+o(i_0^{-3/4})\big)&\quad0\leq\gamma<1\\[9pt]
	\frac{1}{4} \sqrt{\gamma^2-1}\; i_0^{3/2}\,\big(\sqrt{1+2(\de i+\de i_L-\de i_R)}+o(i_0^{-3/4})\big)&\quad\gamma>1\quad.
	\end{cases}
   \label{eq:A pert}
\end{align} 
This formula holds both when the respective sums $\de i_L+\de i_R-\de i$ and $\de i+\de i_L-\de i_R$ vanish and when they are positive and at most of order $O(\sqrt{i_0})$. As a result we have that, when $\de i_L+\de i_R-\de i$,  or $\de i+\de i_L-\de i_R$  respectively, is even the perturbative asymptotics of the square of the $3j$-symbol is 
\begin{align}
	\left(\!\!
\begin{array}{ccc}
\frac{|1-\gamma|}{2} i_0+\de i_L & \frac{(1+\gamma)}{2} i_0+ \de i_R & i_0+\de i \\[4pt]
0 & 0  & 0	
\end{array}
\!\!\right)^2
 &\sim\\
 \sim&\begin{cases}
 \frac{2}{\pi}\frac{1}{\sqrt{1-\gamma^2}}\;\frac{i_0^{-3/2}}{\sqrt{1+2(\de i_L+\de i_R-\de i)}}\;\theta(\de i_L+\de i_R-\de i)&\;\;0\leq\gamma<1\\[9pt]
 \frac{2}{\pi}\frac{1}{\sqrt{\gamma^2-1}}\;\frac{i_0^{-3/2}}{\sqrt{1+2(\de i+\de i_L-\de i_R)}}\;\theta(\de i+\de i_L-\de i_R)&\quad\gamma>1\quad.
 \end{cases}
 	\label{eq:3jperturb}
\end{align}
The theta functions implement the triangular inequality on the fluctuations. In the more general case when $j_1-j_2$ is non-zero but small with respect to the size of the triangle, we have that the fluctuation in $\de j_e$ can be treated perturbatively and, to leading order, the asymptotic expression remains unchanged.

\subsection{Gaussians from factorials}
In this subsection we study the asymptotics of the function $A^i_{i_L i_R}(j_1,j_2)$ which, for $0\leq\gamma<1$, is given by expression (\ref{eq:A=analytic}). The proof in the case $\gamma>1$ goes the same way. In the asymptotic region of interest all the factorials in \eqref{eq:A=analytic} have large argument, therefore Stirling's asymptotic expansion can be used:	
\begin{equation}
	j_0!\;=\;\sqrt{2\pi j_0}\;\;e^{\displaystyle + j_0 (\log j_0\,-1)}\;\;\big(1+\sum_{n=1}^{N}a_n j_0^{-n}\,+\,O(j_0^{-(N+1)})\big)	\quad\qquad \textrm{for all $N>0$,}
\label{eq:stirling}
\end{equation}
where $a_n$ are coefficients which can be computed; for instance $a_1=\frac{1}{12}$. The formula we need is a perturbative expansion of the factorial of $(1+\xi)j_0$ when the parameter $\xi$ is of order $o(1/\sqrt{j_0})$. We have that
\begin{align}
	\big((1+\xi)j_0\big)!=\;\;&\sqrt{2\pi j_0}\;\exp\Big( + j_0 (\log j_0\,-1) + \xi j_0 \log j_0 + j_0 \sum_{k=1}^{\infty}c_k \xi^k\Big)\;\times\\[6pt]
	&\times\;\big(1+\sum_{n=1}^{N} \sum_{m=1}^{M}a_n b_m j_0^{-n}\,\xi^m\;+\,O(j_0^{-(N+\frac{M}{2}+1)})\big)	
\end{align}
where the coefficients $b_m$ and $c_k$ can be computed explicitly. We find that the function $A^i_{i_L i_R}(j_1,j_2)$ has the following asymptotic behavior
\begin{equation}
	A^{i_0+\de i}_{\frac{|1-\gamma|i_0}{2}+\de i_L\,,\,\frac{(1+\gamma)i_0}{2}+\de i_R}(j_0+\de j_1, j_0+\de j_2)\; \sim\; A_0(j_0)\, e^{- H(\de i_L, \de i_R, \de i, \de j_1, \de j_2)}
	\label{eq:asymptotic A}
\end{equation}
where $A_0(j_0)$ is the function evaluated at the background values and $H(\de i_L, \de i_R, \de i, \de j_1, \de j_2)$ is given by
\begin{align}
	H(\de i_L, \de i_R, \de i, \de j_1, \de j_2)=&\;\;\frac{1}{2} (\textrm{arcsinh} \sqrt{3})\, \big(\de i_L +\de i_R-\de i\big)+	\label{eq:H}\\[6pt]
	&+\frac{\sqrt{3}}{2} \frac{(\de i_L)^2}{|1-\gamma| i_0}+\frac{\sqrt{3}}{2} \frac{(\de i_R)^2}{(1+\gamma) i_0}-\frac{\sqrt{3}}{4} \frac{(\de i)^2}{i_0}+\nonumber\\[6pt]
	&-\frac{1}{2}\frac{\de i_L +\de i_R-\de i}{i_0} (\de j_1+\de j_2)\;+\;O(\frac{1}{\sqrt{j_0}})\nonumber
\end{align}
for $0\leq\gamma<1$, while for $\gamma>1$ it is given by
\begin{align}
	H(\de i_L, \de i_R, \de i, \de j_1, \de j_2)=&\;\;\frac{1}{2} (\textrm{arcsinh} \sqrt{3})\, \big(\de i +\de i_L-\de i_R\big)+	\label{eq:H}\\[6pt]
	&+\frac{\sqrt{3}}{2} \frac{(\de i_L)^2}{|1-\gamma| i_0}+\frac{\sqrt{3}}{2} \frac{(\de i_R)^2}{(1+\gamma) i_0}-\frac{\sqrt{3}}{4} \frac{(\de i)^2}{i_0}+\nonumber\\[6pt]
	&-\frac{1}{2}\frac{\de i +\de i_L-\de i_R}{i_0} (\de j_1+\de j_2)\;+\;O(\frac{1}{\sqrt{j_0}})\;.\nonumber
\end{align}

\subsection{Perturbative asymptotics of the fusion coefficients}
Collecting the results of the previous two subsections we find for the fusion coefficients the asymptotic formula
\begin{align}\label{eq:fas}
	f^{i_0+\de i}_{\frac{|1-\gamma|i_0}{2}+\de i_L\,,\,\frac{(1+\gamma)i_0}{2}+\de i_R}(j_0+\de j_e)\;\sim&\;\; f_0(j_0)\;
	\frac{1}{\sqrt{1+2(\de i_L+\de i_R-\de i)}}\;\theta(\de i_L+\de i_R-\de i)\;\times\\[6pt]
	&\times \exp\big(- \textrm{arcsinh}(\sqrt{3})\; (\de i_L+\de i_R - \de i)\big)\;\times\nonumber\\[6pt]
	&\times \exp\big(-\sqrt{3}\frac{(\de i_L)^2}{|1-\gamma|\,i_0}-\sqrt{3}\frac{(\de i_R)^2}{(1+\gamma)\,i_0}+\frac{\sqrt{3}}{2}\frac{(\de i)^2}{i_0}\big)\;\times\nonumber\\[6pt]
	&\times \exp\big(\frac{1}{2}\frac{\de i_L+\de i_R-\de i}{i_0}(\de j_1+\de j_2+\de j_3+\de j_4)\big)\nonumber 	
\end{align}
for $0\leq\gamma<1$, and
\begin{align}\label{eq:fas2}
	f^{i_0+\de i}_{\frac{|1-\gamma|i_0}{2}+\de i_L\,,\,\frac{(1+\gamma)i_0}{2}+\de i_R}(j_0+\de j_e)\;\sim&\;\; f_0(j_0)\;
	\frac{1}{\sqrt{1+2(\de i +\de i_L-\de i_R)}}\;\theta(\de i +\de i_L-\de i_R)\;\times\\[6pt]
	&\times \exp\big(- \textrm{arcsinh}(\sqrt{3})\; (\de i +\de i_L-\de i_R)\big)\;\times\nonumber\\[6pt]
	&\times \exp\big(-\sqrt{3}\frac{(\de i_L)^2}{|1-\gamma|\,i_0}-\sqrt{3}\frac{(\de i_R)^2}{(1+\gamma)\,i_0}+\frac{\sqrt{3}}{2}\frac{(\de i)^2}{i_0}\big)\;\times\nonumber\\[6pt]
	&\times \exp\big(\frac{1}{2}\frac{\de i +\de i_L-\de i_R}{i_0}(\de j_1+\de j_2+\de j_3+\de j_4)\big)\nonumber 	
\end{align}
for $\gamma>1$, where $f_0(j_0)$ is the value of the fusion coefficients at the background configuration. As we will show in next section, this asymptotic expression has an appealing geometrical interpretation and plays a key role in the connection between the semiclassical behavior of the spin foam vertex and simplicial geometries.
\section{Semiclassical behavior}\label{sec:semiclassical}
In \cite{Magliaro:2007ni} the propagation of boundary wave packets was introduced as a way to test the semiclassical behavior of a spinfoam model. In particular, the authors considered an ``initial" state made by the product of four intertwiner wavepackets; this state has the geometrical interpretation of four semiclassical regular tetrahedra in the boundary of a 4-simplex of linear size of order $\sqrt j_0$. Then this state was evolved (numerically) by contraction with the flipped vertex amplitude to give the ``final" state, which in turn is an intertwiner wavepacket. While in \cite{Magliaro:2007ni} only very small $j_0$'s were considered, in \cite{abmp:2008id} we make the same calculation for higher spins both numerically and semi-analitically, and the results are clear: the ``final" state is a semiclassical regular tetrahedron with the same size as the incoming ones. This is exactly what we expect from the classical equations of motion.

The evolution is defined by
\begin{align}
\sum_{i_1\ldots i_5}W(j_0,i_1,\ldots,i_5)\,\psi(i_1,j_0)\ldots\psi(i_5,j_0)\equiv\phi(i_5,j_0),
\label{eq:evolution}
\end{align}
where
\begin{equation}
	\psi(i,j_0)=C(j_0)\,\exp\big(-\frac{\sqrt{3}}{2}\frac{(i-i_0)^2}{i_0}+\mathsf{i} \frac{\pi}{2} (i-i_0)\big)
	\label{eq:initialintertwiner}
\end{equation}
is a semiclassical $SO(3)$ intertwiner (actually its components in the base $|i\rangle$), or a semiclassical tetrahedron, in the equilateral configuration, with $C(j_0)$ a normalization constant, and $W(j_0,i_1,\ldots,i_5)$ is the vertex \eqref{eq:EPRL} with $\gamma=0$ evaluated in the homogeneous spin configuration (the ten spins equal to $j_0$). In \eqref{eq:evolution}, if we want to make the sum over intertwiners, for fixed $j_0$, then we have to evaluate the function $g$ defined as follows:
\begin{equation}
	g(i_L,i_R,j_0)=\sum_i \,f_{i_L\,i_R}^i(j_0) \,\psi(i,j_0)\;.
	\label{eq:g}
\end{equation}
The values of $g$ are the components of an $SO(4)$ intertwiner in the basis $|i_L\rangle |i_R\rangle$, where $|i_L\rangle$ is an intertwiner between four $SU(2)$ irreducible representations of spin $j_0^L\equiv\frac{|1-\gamma|}{2}j_0$, and $|i_R\rangle$ is an intertwiner between representations of spin $j_0^R\equiv\frac{1+\gamma}{2}j_0$. 

We show that EPRL fusion coefficients map $SO(3)$ semiclassical intertwiners into $SU(2)_L\times SU(2)_R$ semiclassical intertwiners. The sum over the intertwiner $i$ of the fusion coefficients times the semiclassical state can be computed explicitly at leading order in a stationary phase approximation, using the asymptotic formula \eqref{eq:fas}\eqref{eq:fas2}. The result is
\begin{align}\label{eq:fasaction}
\sum_i \,f_{i_L\,i_R}^i(j_0) \,\psi(i,j_0)\,\approx\;\alpha_0\; f_0(j_0)\,C(j_0)&\; \times\;\exp\Big(-\sqrt{3}\frac{(i_L-\frac{|1-\gamma|}{2}\,i_0)^2}{|1-\gamma|\,i_0}\pm\mathsf{i}\frac{\pi}{2} (i_L-\textstyle{\frac{|1-\gamma|}{2}}\,i_0)\Big)\;\times\\[6pt]\nonumber
&\;\times\;\exp\Big(-\sqrt{3}\frac{(i_R-\frac{(1+\gamma)}{2}\,i_0)^2}{(1+\gamma)\,i_0}+\mathsf{i}\frac{\pi}{2} (i_R-\textstyle{\frac{(1+\gamma)}{2}}i_0)\Big)
\end{align}
where
\begin{equation}
\alpha_0=\sum_{k\in 2 \mathbb{N}} \frac{e^{-\mathrm{arcsinh}(\sqrt{3})k}}{\sqrt{1+2k}} e^{\mp\mathsf{i}\frac{\pi}{2}k}\simeq 0.97\;;
	\label{eq:alpha0}
\end{equation}
the plus-minus signs both in \eqref{eq:fasaction} and \eqref{eq:alpha0} refer to the two cases $\gamma<1$ (upper sign) and $\gamma>1$ (lower sign).
\begin{figure}[t]
\centering
\begin{minipage}[b]{0.43\textwidth}
\begin{tabular}{c}
\includegraphics[height=0.72\textwidth]{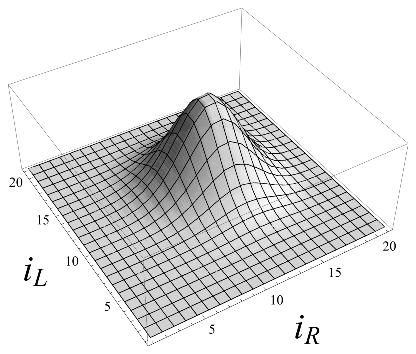}\\[8pt]
(a)
\end{tabular}
\end{minipage}
\begin{minipage}[b]{0.43\textwidth}
\begin{tabular}{c}
\includegraphics[height=0.74\textwidth]{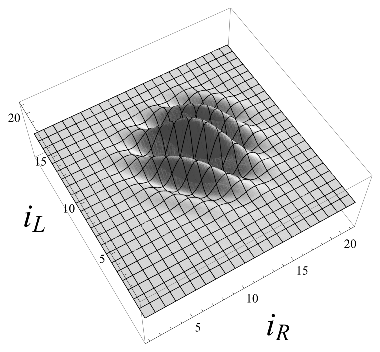}\\
(b)
\end{tabular}
\end{minipage}
\caption{(a) Interpolated plot of the modulus of $g(i_L,i_R,j_0)$ for $j_0=20$ and $\gamma=0$ computed using the exact formula of the fusion coefficients. (b) Top view of the imaginary part.}
\label{fig:plot}
\end{figure}
The r.h.s. of \eqref{eq:fasaction}, besides being a very simple formula for the asymptotical action of the map $f$ on a semiclassical intertwiner, is asymptotically invariant under change of pairing of the virtual spins $i_L$ and $i_R$ (up to a normalization $N$). Recalling that the change of pairing is made by means of $6j$-symbols, we have
\begin{align}\label{eq:invariance}
\sum_{i_L}\sum_{i_R}\sqrt{\mathrm{dim}(i_L)\mathrm{dim}(i_R)}(-1)^{i_L+k_L+i_R+k_R}
&\left\{\!\!
\begin{array}{ccc}
 \frac{|1-\gamma|}{2}j_0& \frac{|1-\gamma|}{2}j_0&i_L\\[3pt]
 \frac{|1-\gamma|}{2}j_0& \frac{|1-\gamma|}{2}j_0&k_L
  \end{array}
	\!\!\right\}\times\\
\times&\left\{\!\!
	\begin{array}{ccc}
  \frac{1+\gamma}{2}j_0& \frac{1+\gamma}{2}j_0&i_R\\[3pt]
  \frac{1+\gamma}{2}j_0& \frac{1+\gamma}{2}j_0&k_R
  \end{array}
	\!\!\right\}
\;g(i_L,i_R)\approx N(j_0)\,g(k_L,k_R,j_0)\;.\nonumber
\end{align}
This result holds because each of the two exponentials in \eqref{eq:fasaction} is of the form
\begin{align}
\exp\big(-\frac{\sqrt3}{2}\frac{(k-k_0)^2}{k_0}\pm\mathsf i\frac{\pi}{2}(k-k_0)\big),
\label{}
\end{align}
which is a semiclassical equilateral tetrahedron with area quantum numbers $k_0$; it follows that $g$ is (asymptotically) an $SO(4)$ semiclassical intertwiner. The formula \eqref{eq:fasaction} can be checked against plots of the exact formula for large $j_0$'s; a particular case is provided in fig.\ref{fig:plot}.

In addition, we can ask whether the inverse map $f^{-1}$ has the same semiclassical property. Remarkably, the answer is positive: $f^{-1}$ maps semiclassical $SO(4)$ intertwiners into semiclassical $SO(3)$ intertwiners. The calculation, not reported here, involves error functions (because of the presence of the theta function) which have to be expanded to leading order in $1/j_0$.

A final remark on our choice for the asymptotic region is needed. The goal we have in mind is to apply the asymptotic formula for the fusion coefficients to the calculation of observables like \eqref{eq:correlation} in the semiclassical regime. If the classical geometry $q$ over which the boundary state is peaked is the geometry of the boundary of a regular 4-simplex, then the sums in \eqref{eq:correlation} are dominated by spins of the form $j_{ab}=j_0+\delta j_{ab}$ and intertwiners of the form $i_a=i_0+\delta i_a$, with $i_0=2 j_0/\sqrt 3$, where the fluctuations must be such that the relative fluctuations $\delta j/j_0,\delta i/j_0$ go to zero in the limit $j_0\rightarrow\infty$. More precisely, the fluctuations are usually chosen to be at most of order $O(\sqrt{j_0})$. This is exactly the region we study in this paper. As to the region in the $(i_L,i_R)$ parameter space, the choice of the background values $\frac{|1-\gamma|}{2}i_0,\frac{1+\gamma}{2}i_0$ and the order of their fluctuations is made {\it a~posteriori} both by numerical investigation and by the form of the asymptotic expansion. It is evident that the previous considerations hold in particular for the function $g$ analyzed in this section.
\section{The case $\gamma=1$}
When $\gamma=1$ we have that $j_L\equiv\frac{|1-\gamma|}{2}j=0$ and we can read from the graph \eqref{eq:f=feyn} that the fusion coefficients vanish unless $i_L=0$. Furthermore it is easy to see that for $\gamma=1$ the fusion coefficients vanish also when $i_R$ is different from $i$. This can be seen, for instance, applying the identity
\begin{equation}
	\parbox[c]{60pt}{\includegraphics{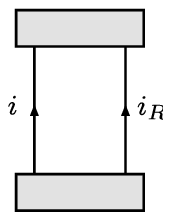}}=\frac{1}{\mathrm{dim}\,i}\,\delta_{i,i_R}\parbox[c]{30pt}{\includegraphics{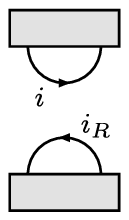}}
	\label{eq:}
\end{equation}
to the graph \eqref{eq:f=feyn} with $i_L=0$. As a result, we have simply
\begin{align}
f^i_{i_L\,i_R}(j_1,j_2,j_3,j_4)=\delta_{i_L,0}\delta_{i_R,i}
\label{eq:}
\end{align}
and the asymptotic analysis is trivial. The previous equation can be also considered as a normalization check; in fact, with the definition \eqref{eq:f=feyn} for the fusion coefficients, the EPRL vertex amplitude \eqref{eq:EPRL} reduces for $\gamma=1$ to the usual $SO(3)$ BF vertex amplitude.
\section{Conclusions}
We summarize our results and give some outlook in a few points.
\begin{itemize}
\item We have shown a simple analytic formula for the LQG fusion coefficients, as defined in the EPRL spinfoam model.
\item We have given a large spin asymptotic formula for the coefficients; specifically, we made a perturbative asymptotic expansion around a background configuration dictated by the kind of boundary state considered.
\item The picture coming out from our analysis is promising: the fusion coefficients not only give nontrivial dynamics to intertwiners at the quantum level, but they seem to behave very well at semiclassical level, in fact they map semiclassical $SO(3)$ tetrahedra into semiclassical $SO(4)$ tetrahedra. This is to us a highly non-trivial property which, in turn, makes the semiclassical analysis of dynamics less obscure. A first application of the asymptotic formula can be found in \cite{abmp:2008id}.
\item Our analysis is a step needed for the study of the full asymptotic expansion of the EPRL vertex, which is part of our work in progress.
\end{itemize}
\section*{Acknowledgments}
\hspace{1.5em} We thank Carlo Rovelli for numerous discussions. E. Bianchi and E. Alesci gratefully acknowledge support by Fondazione Della~Riccia.
\appendix
\section{Properties of $9j$-symbols}\label{app:A}
The $9j$-symbol with two columns with third entry given by the sum of the first two can be written as
\begin{align}
\left\{\!\!
	\begin{array}{ccc}
  a & f & c \\[3pt]
  b & g & d \\[3pt]
  a+b & h & c+d
  \end{array}
	\!\!\right\} =&\;\; (-1)^{f-g+a+b-(c+d)} \;\;\left(\!\!
	\begin{array}{ccc}
  f & g & h \\[3pt]
  a-c & b-d &  -(a+b-(c+d))
  \end{array}
	\!\!\right)\;\;\times
	\label{eq:twodegen}\\[10pt]
	&\hspace{-9em} \times\;\; \sqrt{\frac{(2a)!(2b)!(2c)!(2d)!(a+b+c+d-h)!(a+b+c+d+h+1)!}{(2a+2b+1)!(2c+2d+1)! (a+c-f)!(a+c+f+1)! (b+d-g)! (b+d+g+1)!}} \;.\nonumber
\end{align}
An analogous formula for the $9j$-symbol with two columns with third entry given by the difference of the first two can be obtained from the formula above noting that
\begin{equation}
	\left\{\!\!
	\begin{array}{ccc}
  a & f & c \\[3pt]
  b & g & d \\[3pt]
  b-a & h & d-c
  \end{array}
	\!\right\}=
	 \left\{\!\!
	\begin{array}{ccc}
	b-a & h & d-c\\[3pt]
  a & f & c \\[3pt]
  b & g & d   
  \end{array}
	\!\right\}\;,
	\label{eq:}
\end{equation}
so we are in the previous case.\\
The $3j$-symbol with vanishing magnetic numbers has the simple expression
\begin{equation}
	\left(\!\!
	\begin{array}{ccc}
  a & b & c \\[3pt]
  0 & 0 & 0
  \end{array}
	\!\!\right) = (-1)^{a-b}  \pi^{1/4} \frac{2^{\frac{a+b-c-1}{2}}}{(\frac{c-a-b-1}{2})! \sqrt{(a+b-c)!}}\;\; \sqrt{\frac{(\frac{c+a-b-1}{2})! (\frac{c-a+b-1}{2})! (\frac{a+b+c}{2})!}{
	 (\frac{c+a-b}{2})! (\frac{c-a+b}{2})! (\frac{a+b+c+1}{2})!}}\;.
	\label{eq:}
\end{equation}
These formula can be derived from \cite{YutsinLevinson:1962,Varsh:1988}.
\section{Regge asymptotic formula for $3j$-symbols}\label{app:B}
The asymptotic formula of $3j$-symbols for large spins $a,b,c$ and admitted magnetic numbers, i.e. $m_a+m_b+m_c=0$, given by G. Ponzano and T. Regge in \cite{PonzanoRegge:1968} is
\begin{equation}
	\left(\!\!
	\begin{array}{ccc}
  a & b & c \\[3pt]
  m_a & m_b & m_c
  \end{array}
	\!\!\right) \sim \frac{(-1)^{a+b-c+1}}{\sqrt{2 \pi A}} \cos\Big( (a+\frac{1}{2})\theta_a+(b+\frac{1}{2})\theta_b+(c+\frac{1}{2})\theta_c+ m_a \phi_a -m_b \phi_b +\frac{\pi}{4}\Big)
 	\label{eq:3jRegge}
\end{equation}
with
\begin{align}
&\theta_a=\frac{\arccos \Big(2(a+\onehalf)^2 m_c+m_a\big((c+\onehalf)^2+(a+\onehalf)^2-(b+\onehalf)^2\big)\Big)}
                     {\sqrt{\big((a+\onehalf)^2-m_a^2\big)\Big(4(c+\onehalf)^2(a+\onehalf)^2-\big((c+\onehalf)^2+
                      (a+\onehalf)^2-(b+\onehalf)^2\big)^2\Big)}}\\
&\phi_a=\arccos\left(\onehalf\frac{(a+\onehalf)^2-(b+\onehalf)^2-(c+\onehalf)^2-2 m_b m_c}
          {\sqrt{\big((b+\onehalf)^2-m_b^2\big)\big((c+\onehalf)^2-m_c^2\big)}}\right)\\
&A=\sqrt{-\frac{1}{16}\det \left(
\begin{array}{cccc}0&(a+\onehalf)^2-m_a^2&(b+\onehalf)^2-m_b^2&1\\[3pt]
(a+\onehalf)^2-m_a^2&0&(c+\onehalf)^2-m_c^2&1\\[3pt]
(b+\onehalf)^2-m_b^2&(c+\onehalf)^2-m_c^2&0&1\\[3pt]
1&1&1&0
\end{array}\right)}
\label{}
\end{align}
and $\theta_b,\theta_c,\phi_b$ are obtained by cyclic permutations of $(a,b,c)$.

\providecommand{\href}[2]{#2}\begingroup\raggedright\endgroup

\end{document}